\documentclass[conference]{IEEEtran}
\IEEEoverridecommandlockouts

\usepackage{amsmath,amssymb,amsfonts,amsthm}
\usepackage{mathtools}
\usepackage{booktabs}
\usepackage{graphicx}
\usepackage{algorithm}
\usepackage{algpseudocode}
\usepackage{cite}
\usepackage{xurl}
\usepackage{hyperref}

\usepackage{comment}

\newtheorem{theorem}{Theorem}
\newtheorem{proposition}{Proposition}
\newtheorem{corollary}{Corollary}
\theoremstyle{remark}
\newtheorem{remark}{Remark}
\theoremstyle{definition}
\newtheorem{definition}{Definition}

\newcommand{\F}{\mathbb{F}}
\newcommand{\E}{\mathbb{E}}
\renewcommand{\Pr}{\mathbb{P}}
\newcommand{\KL}{D}
\DeclareMathOperator{\Bin}{Bin}
\DeclareMathOperator{\Poi}{Poi}

\newcommand{\floor}[1]{\lfloor #1\rfloor}

\begin{document}

\title{Concrete Cryptanalysis of the Permuted Reed--Solomon Pseudorandom Code for LLM Watermarking}

\author{\IEEEauthorblockN{Shuiyin Liu}
\IEEEauthorblockA{\textit{Cyber Security Research and Innovation Centre} \\
\textit{Holmes Institute}\\
Melbourne, VIC 3000, Australia \\
Email: SLiu@Holmes.edu.au}
\and
\IEEEauthorblockN{Amin Sakzad}
\IEEEauthorblockA{\textit{Department of Software Systems and Cybersecurity} \\
\textit{FIT, Monash University}\\
Melbourne, VIC 3800, Australia \\
Email: Amin.Sakzad@monash.edu}}

\maketitle

\begin{abstract}
Pseudorandom error-correcting codes (PRCs) give undetectable and robust watermarks for large language models (LLMs).
To our knowledge, the permuted Reed–Solomon code (PRSC) is the most robust PRC against worst-case edits, and its security rests on an asymptotic conjecture that has not been tested at deployable sizes.
We give a keyless distinguisher for PRSC that counts the symbols at which two responses agree, a count that the secret permutations leave unchanged.
We derive the exact distribution of the agreement count and the resulting sample cost $T\approx6/\sqrt{\KL}$, where $\KL$ is the information per pair of responses.
The cost $T$ grows as $2^{\Theta(k\log k)}$ in the message dimension $k$, which is superpolynomial in the code length $n$, so it does not contradict the conjecture, which requires $T=\mathrm{poly}(n)$.
Expressed in tokens, the cost depends on the language model only through its mean bit-flip probability $\bar p$ and its embedded bits per token $\kappa$. Under an i.i.d.\ symbol model, we show that no dimension $k\ge2$ is decodable within the Guruswami--Sudan radius once $\bar p\ge1-1/\sqrt2$.
With measured GPT-2 and Llama-2-7B statistics at gate $\tau=0.2$, this translates to the fact that a $300$-word response gives a cost between $2^{29}$ and $2^{58}$ (extra channel noise $\eta=0.1$ or $0.2$), and $2^{128}$ needs about 1000--1600 words for GPT-2 and 3500--4400 for Llama-2-7B.
Our end-to-end experiments on both models, at code length $17$, agree with the predicted cost within about 0.3 bit in most settings.
\end{abstract}

\section{Introduction}
A \emph{pseudorandom error-correcting code} (PRC) is a keyed code whose codewords are computationally indistinguishable
from uniform strings without the key, yet which the key holder can decode after a constant fraction of the symbols have been
corrupted \cite{ChristGunn24,Christ26}.
PRCs were introduced to watermark large language models
(LLMs): the random seed of the generator is replaced by a PRC codeword, which leaves the
output distribution computationally unchanged (\emph{undetectability}), and the key holder detects the watermark by recovering the seed from
the text and decoding, despite sampling noise and user edits \cite{ChristGunn24,Christ26}.

The first PRC is built from LDPC codes and rests on the learning-parity-with-noise (LPN) and planted-XOR assumptions \cite{ChristGunn24}.
Later work widened the set of assumptions. Golowich and Moitra~\cite{GM24} obtain a secret-key PRC from an $O(\log n)$-local weak
pseudorandom function (wPRF), and use it to obtain edit-distance-robust watermarks.
Ghentiyala and Guruswami~\cite{GG25} obtain a public-key PRC with $o(1)$ distinguishing advantage from the planted-hyperloop assumption of
\cite{BKR23} together with Goldreich's local pseudorandom generator (PRG), and also weaken the planted-XOR assumption behind the LDPC construction.
In all three, the decoder inspects a sparse $\Theta(\log n)$-size planted structure, which admits $n^{\Theta(\log n)}$-time distinguishers \cite{Christ26}.
Christ, Golowich, Gunn, Moitra and Wichs~\cite{CGGMW26}  removed this limitation with the \emph{permuted-codes} construction: a codeword of a code with large dual distance is scrambled by a secret coordinate permutation $\sigma$ and symbol permutations $\pi_1,\dots,\pi_n$, and then passed through a substitution channel that independently replaces each symbol with a uniformly random element of $\mathbb{F}_q$ with probability $1-\rho$. Thus, the decoder must exploit the structure of the whole noisy codeword. Its security rests on the permuted-puzzles conjecture \cite{BIPW17,CHR17}.
 Instantiated with Reed–Solomon (RS) codes, it gives the permuted RS code (PRSC). The permuted RS conjecture states that PRSC is pseudorandom against subexponential-time distinguishers  \cite{CGGMW26}. Variants built from folded RS codes and from binarized RS codes tolerate a constant fraction of worst-case edits (insertions, deletions and substitutions) in the watermarked text  \cite{CGGMW26}. To our knowledge PRSC is the most robust PRC, meaning that its variants still decode under worst-case edits and not only random substitutions.

For the LDPC construction, Wang \emph{et al.}\ gave the first systematic cryptanalysis \cite{WWC25}: a partial key-recovery attack, a
weak-key distinguisher and a noise-overlay removal attack.
They detect a watermark with about $2^{22}$ operations on DeepSeek and Stable Diffusion, and show that $2^{128}$ security needs a code
length far beyond the output length of current LLMs \cite{WWC25}.
No comparable analysis exists for PRSC, which has no sparse parity check to recover; attacks on such assumptions are an open problem \cite{Christ26}.

PRSC's pseudorandomness is asymptotic: for an RS code of length $n$ and dimension $k$ (the number of message symbols; any $k$ symbols determine a codeword) with $k=n^{\Omega(1)}$, it is conjectured to hold against $T=\mathrm{poly}(n)$ samples, when each sample is independently corrupted by a constant-rate substitution channel. Here, one sample is one watermarked response of \(n\) symbols, in which each symbol is retained with probability \(\rho\) and otherwise replaced uniformly at random over \(\mathbb F_q\). A deployed watermark has concrete parameters: a few hundred to a few thousand tokens per response and a noise level set by the language model.
We study PRSC at these concrete sizes in two steps.
First, we give an explicit distinguisher: a keyless test that counts the coordinates at which two watermarked responses agree, a count that is invariant under the secret permutations. We derive its exact distribution and its sample cost, which grows as $2^{\Theta(k\log k)}$. This is superpolynomial in $n$, so our distinguisher does not contradict the conjecture, which concerns $\mathrm{poly}(n)$ samples.
Second, we use this analysis to answer a practical question:
\begin{quote}\emph{For a given LLM, token budget and noise level, how many watermarked responses does this distinguisher need, and is this above or below $2^{128}$?}
\end{quote}
The response length required to achieve $2^{\lambda_{\rm s}}$ security ($\lambda_{\rm s}$ is the security parameter in bits) grows only polynomially in $\lambda_{\rm s}$, specifically as $\lambda_{\rm s}^{1/(1-2a)}$ with $a=\log_2\frac1{1-\bar p}$. The degree $1/(1-2a)$ is about $2$ when the mean bit-flip probability $\bar p$ of the embedded decisions is about $0.16$, as for GPT-2 and Llama-2-7B at gate $\tau=0.2$. A model with $\bar p\ge1-1/\sqrt2\approx0.29$ admits no decodable dimension $k\ge2$ at all. Our contributions are as follows. All symbols are defined where they first appear and collected in Table~\ref{tab:notation}.

\begin{itemize}
\item \textbf{A distinguisher for PRSC.} We construct a keyless distinguisher that counts the coordinates at which two responses agree, a quantity that the secret key $(\sigma,\pi_1,\dots,\pi_n)$ (a coordinate permutation and $n$ symbol permutations) leaves unchanged. We derive the exact distribution of this count from the root count of the difference polynomial (Theorems~\ref{thm:law}--\ref{thm:chi}). We show that the distinguisher needs $T\approx6/\sqrt\KL$ watermarked responses, where $\KL$ is the divergence (in nats) between the agreement count of two watermarked responses and that of two uniform words, where $\KL\approx(\vartheta\rho^4)^k/(2\,k!)$ for code length $n$, dimension $k$, load $\vartheta=n/q$ and symbol survival probability $\rho$ (Proposition~\ref{prop:cost}). At the largest dimension the decoder can handle, $k=\lfloor n\rho^2\rfloor$, we obtain $\log_2T\approx n\rho^2[\tfrac12\log_2(n/e)+\log_2(1/\rho)]$ (Corollary~\ref{cor:nogo}). This cost is superpolynomial in the security parameter, so the distinguisher does not refute the conjecture. It shows instead how far the constants of PRSC are from $2^{128}$ at deployable sizes.
\item \textbf{Cost for LLM watermarks.} We express the cost through the response length $B$ in tokens, the gate $\tau$ (the minimum margin for a token decision to carry a code bit), and two statistics of the language model: the number $\kappa$ of embedded bits (binary decisions) per token and the mean bit-flip probability $\bar p$ of an embedded decision, that is, the chance that sampling flips the bit the code wants (Theorem~\ref{thm:token}). We prove that, under the i.i.d.\ symbol model, no $k\ge2$ is decodable within the Guruswami--Sudan radius once $\bar p\ge1-1/\sqrt2$ (Corollary~\ref{cor:tokens}). At $\tau=0.2$ we find that a response needs about 1000--1600 words (GPT-2 \cite{GPT2}) or 3500--4400 words (Llama-2-7B \cite{Llama2}) to reach $2^{128}$ against this distinguisher. 
\item \textbf{Validation.} We check our asymptotically derived closed form formulas numerically against the exact distributions (Tables~\ref{tab:series}--\ref{tab:token}). For responses of 300 words at gate $\tau=0.2$, the cost in Table~\ref{tab:llm} is between $2^{29}$ and $2^{58}$ for GPT-2 and Llama-2-7B (extra noise $\eta=0.1$ or $0.2$), below the $2^{128}$ target. We then run the attack on GPT-2 and Llama-2-7B watermarks at code length $n=17$. It separates watermarked from control responses with $T=1000$ responses for GPT-2 ($k=2,3$; AUC $1.00$) and with $T=500$ for Llama-2-7B (AUC $\ge0.98$), and the measured cost agrees with the cost predicted by our theory (Theorem~\ref{thm:law} and Proposition~\ref{prop:cost}) within about $0.3$ bit in most settings. 

\end{itemize}

The rest of the paper is organized as follows. Section~\ref{sec:channel} models the LLM channel. Section~\ref{sec:main} defines the pair test and states the main results, where proofs are given in the Appendix. Section~\ref{sec:val} validates them numerically.

\section{Background and Notation}
In this section, we first provide a short background on PRC based on permuted RS codes. We then describe these in the context of LLMs. Table~\ref{tab:notation} summarises the notation. The pair-test quantities ($A$, $P_{\rm WM}$, $\KL$, $T$) are introduced in Sec.~\ref{sec:pair}.

\begin{table}[!t]
\centering
\caption{Notation}
\label{tab:notation}
\small
\begin{tabular}{@{}lp{0.64\columnwidth}@{}}
\toprule
$\F_q$ & finite field with $q$ elements\\ 
$\mathrm{Sym}(\F_q)$ & is the group of all permutations of $\F_q$\\
$S_n$ & $\mathrm{Sym}(\{1,\dots,n\})$\\
$x\leftarrow S$ & $x$ is drawn uniformly at random from set $S$\\
$\mathrm{Unif}(S)$ & uniform distribution on $S$ \\
$\mathrm{Ber}(p)$ & is Bernoulli with $\Pr[1]=p$\\
$\Bin(n,p)$ & binomial distr. $\Pr[x]=\binom nxp^x(1-p)^{n-x}$\\ $\Poi(\vartheta)$ & Poisson with mean $\vartheta$ and $\Pr[x]=e^{-\vartheta}\vartheta^x/x!$\\
$\Pr[\cdot],\ \E[\cdot]$ & probability and expectation\\
$\approx,\ \ll,\ \lesssim,\ \gtrsim$ & approximately equal (the accuracy is stated where used); much smaller than; at most, and at least, up to a constant factor\\
$O(\cdot),\ \Theta(\cdot)$ & bounded by, and of the same order as, a constant times the argument\\
$\mathrm{Dist}$ & a distinguisher: a (computationally bounded) algorithm that outputs $1$ or $0$ on its input samples\\
$Y^{(t)},\ U^{(t)}$ & $t$-th watermarked sample (a response) and $t$-th uniform sample, each in $\F_q^n$\\
\midrule
$n,\,q,\,k,\,L$ & RS length, alphabet size (a prime power), dimension; $q=2^L$ in the LLM watermark ($L$ bits per symbol)\\
$K=(\sigma,\pi_j)$ & secret index / alphabet permutations\\
$\hat p_i,\ m_i,\ p_i$ & $\Pr[\text{bit}=1]$; $\min(\hat p_i,1-\hat p_i)$; bit-flip probability $\tfrac12-m_i$\\
$\bar p$ & average bit-flip probability over embedded decisions\\
$V,\ \ell_V$ & vocabulary size; binary decisions per token, $\ell_V=\lceil\log_2V\rceil$\\
$\tau,\ \kappa(\tau)$ & gate threshold; embedded bits per token ($\kappa\le\ell_V$)\\
$B$ & response length in tokens ($n\approx B\kappa/L$)\\
$\eta$ & extra channel substitution rate (edits)\\
$\rho=1-\eta_{\rm eff}$ & symbol survival, Eq.~\eqref{eq:rho}; $s=\rho^2$ (both samples intact)\\
$\vartheta=n/q$ & load of the evaluation set ($\approx1$ in PRSC)\\
$\lambda_{\rm s}$ & security parameter (bits); PRSC sets $n\approx\lambda_{\rm s}$\\
$A$ & pair agreement count\\
$Q(z)$ & Gaussian tail $1-\Phi(z)$\\
$\varsigma$ & standard deviation of the per-pair log-likelihood ratio (Rem.~\ref{rem:time})\\
$\KL(P\Vert Q)$ & Kullback--Leibler divergence $\sum_xP(x)\ln\frac{P(x)}{Q(x)}$ (nats); $\KL=\KL(P_{\rm WM}\Vert P_0)$ is the information per pair, Eq.~\eqref{eq:Ddef}\\
$\chi^2(P\Vert Q)$ & $\chi^2$ divergence $\sum_x\frac{(P(x)-Q(x))^2}{Q(x)}$, defined when $Q>0$; $\KL(P\Vert Q)\le\chi^2(P\Vert Q)$\\
$P_{\rm WM},P_0$ & distribution of $A$: watermarked pair with distinct codewords ($g\ne0$), Thm.~\ref{thm:law}(c) / uniform pair $\Bin(n,1/q)$\\
$T$ & number of watermarked responses (samples)\\
$k_{\rm dec}$ & Guruswami--Sudan (GS) decoding limit, Eq.~\eqref{eq:kdec}\\
$a$ & $\log_2\frac1{1-\bar p}$ (decay exponent of the symbol survival)\\
$L_{\min}$ & smallest $L$ with $B\kappa/L\le2^L+1$ (no unused bits)\\
$p^\ast$ & no-go threshold $1-1/\sqrt2\approx0.293$\\
\bottomrule
\end{tabular}
\end{table}

\subsection{PRCs and the permuted RS conjecture}
Let $q$ be a prime power (in the LLM watermark $q=2^L$, so that each symbol carries $L$ bits) and let $\mathcal E\subset \F_q\cup\{\infty\}$ be a set of $n$ evaluation points (the \emph{evaluation set})
($n\le q$, or $n\le q+1$ for extended RS). The RS code is $C=\{(f(\alpha))_{\alpha\in\mathcal E}: f\in\F_q[x]_{<k}\}$ (the value at $\infty$ is the
coefficient of $x^{k-1}$), which is a maximum distance separable (MDS) code: any $k$ evaluations determine $f$, so two distinct codewords agree in at most $k-1$ positions.

\begin{definition}[Permuted code samples \cite{CGGMW26}]
The secret key is $K=(\sigma,\pi_1,\dots,\pi_n)$, with $\sigma\in S_n$ an index permutation and $\pi_j\in\mathrm{Sym}(\F_q)$ alphabet
permutations. One sample is $\hat c_j=\pi_j(c_{\sigma(j)})$ with $c\leftarrow C$ uniform, followed by the substitution channel
$Y_j=\hat c_j$ with probability $\rho$ and $Y_j\sim\mathrm{Unif}(\F_q)$ otherwise. The $T$ samples of one key are $Y^{(1)},\dots,Y^{(T)}$.
\end{definition}
\emph{Conjecture} (PRSC, \cite{CGGMW26} Conj.~3.2): For a given security parameter $\lambda_s$, let $n=\lambda_{\rm s}$, $q=\lambda_{\rm s}-1$, $k=\lambda_{\rm s}^{1/5}$ and $\rho$ be a constant noise. With these parameters, any $T=\mathrm{poly}(\lambda_{\rm s})$
samples are computationally indistinguishable from uniform. The distinguisher's advantage is
$|\Pr[\mathrm{Dist}(Y^{(1..T)})=1]-\Pr[\mathrm{Dist}(\mathrm{Unif}^{(1..T)})=1]|$.
Dropping any one of $\sigma$, $\pi_j$ or the noise breaks the conjecture. For example, without the $\pi_j$ an algebraic attack finds a low-degree polynomial that vanishes on the sample columns \cite[Sec.~3.1]{CGGMW26}, while with $\pi_j$ such relations are destroyed.

\subsection{From a PRC to an LLM watermark}\label{sec:channel}

Let $V$ be the vocabulary size ($50,257$ for GPT-2, $32,000$ for Llama-2-7B). An LLM emits a token as a sequence of $\ell_V=\lceil\log_2 V\rceil$ binary decisions along a fixed binary code of the vocabulary
($\ell_V=16$ for GPT-2, $15$ for Llama-2-7B). At decision $i$ the model gives $\hat p_i=\Pr[\text{bit}=1\mid\text{prefix}]$ and we write
$m_i=\min(\hat p_i,1-\hat p_i)$. To embed a (uniform) codeword bit $c_i$, the watermark draws the token bit $b_i$ from
$\mathrm{Ber}(\hat p_i-(-1)^{c_i} m_i)$ \cite{ChristGunn24,CGGMW26}. This is the optimal coupling: if $c_i$ is uniform, $b_i$ has exactly the model's
distribution (undetectability), and $b_i$ differs from $c_i$ with probability
\begin{equation}
p_i \;=\; \tfrac12-m_i \;=\;\big|\hat p_i-\tfrac12\big|\;\in[0,\tfrac12].
\label{eq:pi}
\end{equation}
So $p_i$ \emph{depends on the decision}: $p_i=0$ if $\hat p_i=\tfrac12$ (noiseless), $p_i=\tfrac12$ if $\hat p_i\in\{0,1\}$ (fully erased), and $p_i=|\hat p_i-\tfrac12|$ in between. Here $p_i$ averages over the uniform bit $c_i$.

\textbf{Channel.} We write $\bar p=\E[p_i]$ for the \emph{average bit-flip probability} over the embedded decisions, and \emph{model the embedding as a binary symmetric channel (BSC) with crossover probability $\bar p$}. We use it to estimate symbol survival probability.

\textbf{Gating.} Gating keeps only decisions with $p_i\le\tfrac12-\tau$. Without gating, many decisions are nearly erased: at $\tau=0$, $65\%$ (Llama-2-7B) and $36\%$ (GPT-2) of the measured decisions have $p_i\ge0.49$ (Sec.~\ref{sec:etaeff}). Gating lowers $p_i\le\tfrac12-\tau$, but fewer decisions
qualify: $\kappa(\tau)\le\ell_V$ denotes the average number of embedded bits per token after gating ($\kappa=\ell_V$ if no decision is gated out, e.g.\ $\tau=0$). A response of $B$ tokens
therefore carries about $B \cdot \kappa(\tau)$ bits.

\textbf{Symbols.} Each RS symbol is written as $L=\log_2 q$ consecutive embedded bits, so
$n\approx B\cdot \kappa(\tau)/L$. A symbol survives only if all its $L$ bits do. With an additional channel substitution rate $\eta$
(edits, paraphrase), the \emph{effective symbol survival probability} is
\begin{equation}
\rho \;:=\;1-\eta_{\rm eff}\;=\;(1-\eta)\;\E\!\Big[\prod_{i=1}^{L}(1-p_i)\Big]\;\approx\;(1-\eta)(1-\bar p)^{L}.
\label{eq:rho}
\end{equation}
A corrupted symbol is modelled as uniform on $\F_q$, which is the substitution channel of the conjecture. A real symbol with only some bits flipped can still coincide between two samples, so this is an approximation; the empirical control in Sec.~\ref{sec:val} absorbs it.

\textbf{Decoding limit.} The Guruswami--Sudan list decoder \cite{GS99} recovers a codeword that agrees with the received word in
$t>\sqrt{(k-1)n}$ positions. Since $t\approx n\rho$, the exact condition is $k-1<n\rho^2$, i.e.\ $k\le\lceil n\rho^2\rceil$. We use its large-$n$ form (the two differ by at most one unit), so a polynomial-time detector needs
\begin{equation}
k\;\le\;k_{\rm dec}\;:=\;\floor{n\rho^2}.
\label{eq:kdec}
\end{equation}
An exhaustive, exponential-time detector can use larger $k$.

\section{Main Results}\label{sec:main}
In this section, we first present our pair-agreement test algorithm. We then provide theoretical results calculating asymptotic number of samples required to run the test and as well as asymptotic time and token complexities. All proofs are provided in Appendix.
\subsection{The pair-agreement test (our attack)}\label{sec:pair}

\textbf{The test (see Algorithm~\ref{alg:pair}).} Let $Y,Y'\in\F_q^n$ be two responses generated under the same key $K$, and let $A=\#\{j: Y_j=Y'_j\}$ be the number of positions where they agree. Because $\sigma$ and every $\pi_j$ are bijections, \emph{$A$ does not depend on the key} (Theorem~\ref{thm:law}(c)), so the attacker needs no key knowledge. 

Let $P_0(x)=\Pr[\Bin(n,1/q)=x]$ be the distribution of $A$ for two uniform words. Let $P_{\rm WM}(x)$ be the distribution of $A$ for two independent watermarked responses under the same key whose underlying codewords are distinct, i.e., given $g=f-f'\ne0$; this is the $P_{\rm WM}$ of Theorem~\ref{thm:law}(c).
The attacker receives $T$ responses  $Y^{(1)},\dots,Y^{(T)}\in\F_q^n$ that are either all watermarked under one unknown key or all uniform. It computes $A_{ij}$ for each of the $\binom T2$ pairs and sums the log-likelihood ratio $\ln(P_{\rm WM}(A)/P_0(A))$ over the pairs (Proposition~\ref{prop:cost}).

The signal is easy to see. For two uniform words, $A\sim\Bin(n,1/q)$. For RS, two distinct noise-free codewords agree in at most $k-1$ positions, so the counts $A\ge k$ are missing from the watermarked distribution. Noise weakens this signal: a shared codeword symbol is seen as an agreement only if both copies survive (probability $\rho^2$), and replaced symbols agree by chance, so counts $A\ge k$ reappear. Theorem~\ref{thm:chi} measures the remaining signal by the information $\KL$ per pair.

Algorithm~\ref{alg:pair} presents the test in full. It first builds the log-likelihood-ratio table $h(x)=\ln(P_{\rm WM}(x)/P_0(x))$ from Theorem~\ref{thm:law}. It then sums $h$ over all pairs of responses and compares the sum with a threshold.
\begin{algorithm}[t]
\caption{Pair-agreement distinguisher (keyless)}\label{alg:pair}
\small
\begin{algorithmic}[1]
\Require Responses $Y^{(1)},\dots,Y^{(T)}\in\F_q^n$; parameters $n,q,k$ and symbol survival $\rho$
\Ensure ``watermarked'' or ``uniform''
\State $P_0\gets\Bin(n,1/q)$; $P_{\rm WM}\gets$ distribution from Theorem~\ref{thm:law}(c)
\State $h(x)\gets\ln\big(P_{\rm WM}(x)/P_0(x)\big)$ for $x=0,\dots,n$
\State $N\gets\binom T2$; $\ \theta\gets\tfrac N2\big(\sum_xP_{\rm WM}(x)h(x)+\sum_xP_0(x)h(x)\big)$
\State $W\gets0$
\For{each pair $1\le i<j\le T$}
\State $A_{ij}\gets\#\{\ell:\ Y^{(i)}_\ell=Y^{(j)}_\ell\}$
\State $W\gets W+h(A_{ij})$
\EndFor
\If{$W>\theta$} \State \Return ``watermarked'' \Else \State \Return ``uniform'' \EndIf
\end{algorithmic}
\end{algorithm}
 
\textbf{Information per pair.} 
The identical-codeword event has probability $q^{-k}$ and is handled in Theorem~\ref{thm:law}(a) and in Sec.~\ref{sec:e2e}. The quantity that governs every cost result below is the Kullback--Leibler divergence
\begin{equation}
\begin{split}
\KL&:=\KL\big(P_{\rm WM}\Vert P_0\big)=\sum_{x=0}^{n}P_{\rm WM}(x)\ln\frac{P_{\rm WM}(x)}{P_0(x)}\\
&=\E_{\rm WM}\!\left[\ln\frac{P_{\rm WM}(A)}{P_0(A)}\right]\ \text{(nats)}.
\end{split}
\label{eq:Ddef}
\end{equation}
It is the average evidence, in nats, that one pair of responses provides in favour of ``watermarked'' over ``uniform''. It is $0$ exactly when $P_{\rm WM}=P_0$, and the evidence adds up over pairs. A small $\KL$ therefore means a secure code and a large $\KL$ a cheap attack. Theorem~\ref{thm:chi} computes $\KL$, and Proposition~\ref{prop:cost} turns it into the number of responses $T$ needed.

\textbf{Cost and threat model.} We call $T$ the \emph{cost}: the number of watermarked responses the attacker needs under one key. The running time is about $nT^2/2$ (Remark~\ref{rem:time}). The \emph{provider} is the party that deploys the PRSC watermark in a language model and holds the key. The attacker has no key, but knows which decisions carry bits (aligned positions). This is the setting of the experiments in Sec.~\ref{sec:e2e}.

\subsection{Main theorems}

The results of this subsection that involve only the code and the symbol noise (Theorems~\ref{thm:law}--\ref{thm:chi}, Proposition~\ref{prop:cost} and Corollary~\ref{cor:nogo}) hold for any prime power $q$. Theorem~\ref{thm:token} and Corollary~\ref{cor:tokens} use $q=2^L$, because they map $L$ LLM bits to one symbol. Theorem~\ref{thm:law} gives the exact distribution of the agreement count $A$ of two responses. Theorem~\ref{thm:chi} turns it into the information that one pair of responses carries about the watermark, and Proposition~\ref{prop:cost} converts that information into the number of responses an attacker needs. Corollary~\ref{cor:nogo} and Theorem~\ref{thm:token} then evaluate this cost at the decoding limit and for LLM responses of a given length. Section~\ref{sec:val} checks the formulas numerically.
 
Recall that $\mathcal E$ is the set of $n$ evaluation points and that each noise-free codeword is $(f(\alpha))_{\alpha\in\mathcal E}$ for a message polynomial $f$ of degree less than $k$. Throughout, $s=\rho^2$ is the probability that a given symbol survives the noise in both of two responses.

\begin{theorem}[Exact pair distribution]\label{thm:law}
Let $Y$ and $Y'$ be two independent responses under the same key $K$, and let $A$ be the number of positions where they agree. Each response is generated as follows. Draw a uniform message polynomial ($f$ for $Y$, $f'$ for $Y'$) of degree less than $k$. Apply the key $K$, which permutes the positions and the alphabet at each position. Then keep each symbol with probability $\rho$ and otherwise replace it by a uniform symbol of $\F_q$, independently. Let $s=\rho^2$ be the probability that both symbols at a position are kept, and $g=f-f'$. The roots of $g$ in $\mathcal E$ are exactly the evaluation points at which the two noise-free codewords agree. Let $R$ be the number of these roots. Then:
\begin{enumerate}
\item[(a)] \emph{Root count.} $\Pr[g=0]=q^{-k}$, in which case the two codewords are identical. Given $g\neq0$, we have $R\le k-1$ and
\[
\Pr[R=r\mid g\ne0]=\sum_{m=r}^{k-1}(-1)^{m-r}\binom mr\binom nm\,\psi_m,
\]
where $\psi_m=\frac{q^{k-m}-1}{q^k-1}$ is the probability that $g$ vanishes on a fixed set of $m$ evaluation points, given $g\ne0$.
\item[(b)] \emph{Agreement count.} Given $g\ne0$ and $R=r$, $A$ is a sum of two independent binomials,
$A\sim\Bin(r,\gamma_1)+\Bin(n-r,\gamma_0)$, where $\gamma_1=s+\tfrac{1-s}{q}$ and $\gamma_0=\tfrac{1-s}{q}$.
If $g=0$, then $A\sim\Bin(n,\gamma_1)$, which is the case $r=n$.
\item[(c)] \emph{Distribution of $A$.} Write $\beta(i;m,\gamma)=\binom mi\gamma^i(1-\gamma)^{m-i}$ for the binomial probability. Averaging over $f,f'$ and the noise,
\[
\Pr[A=x]=q^{-k}\beta(x;n,\gamma_1)+(1-q^{-k})\,P_{\rm WM}(x),
\]
where, for $x=0,\dots,n$, the distribution given $g\ne0$, which we denote $P_{\rm WM}$, is
\[
\begin{aligned}
P_{\rm WM}(x)=\sum_{r=0}^{k-1}&\Pr[R=r\mid g\ne0]\\
&\cdot\sum_{i=0}^{r}\beta(i;r,\gamma_1)\,\beta(x-i;n-r,\gamma_0).
\end{aligned}
\]
\item[(d)] \emph{Key independence.} The distribution of $A$ is the same for every key $K$. For two uniform words, $A\sim\Bin(n,1/q)$.
\end{enumerate}
\end{theorem}

We like to know ``how much does one pair of watermarked responses reveal?'' We measure this with the $\chi^2$ divergence between the agreement-count distribution $P_{\rm WM}$ of a watermarked pair and the distribution $P_0$ of a uniform pair, that is
\begin{equation}
\chi^2(P_{\rm WM}\Vert P_0)=\sum_{x=0}^{n}\frac{\big(P_{\rm WM}(x)-P_0(x)\big)^2}{P_0(x)}. \label{chi_D}
\end{equation}
It is zero exactly when the two distributions coincide, and it is convenient here because it has a closed form as a series. The Kullback--Leibler divergence $\KL$, is bounded by $\chi^2$ and is about $\tfrac12\chi^2$ when the two distributions are close.

\begin{theorem}[Information per pair]\label{thm:chi}
Fix $k$ and $\rho$, and let $q\to\infty$ with $n/q\to\vartheta\in(0,1]$. Let $P_{\rm WM}$ be the distribution of $A$ given $g\ne0$ from Theorem~\ref{thm:law}(c), and $P_0=\Bin(n,1/q)$ the distribution for two uniform words; $P_0$ tends to the Poisson distribution $\Poi(\vartheta)(x)=e^{-\vartheta}\vartheta^x/x!$. We also have
\begin{equation}
\chi^2(P_{\rm WM}\Vert P_0)\;\longrightarrow\;\sum_{m\ge k}\binom{m-1}{k-1}^{2}\frac{(\vartheta\rho^{4})^{m}}{m!}.
\label{eq:chi}
\end{equation}
In addition, $\KL(P_{\rm WM}\Vert P_0)\le\chi^2(P_{\rm WM}\Vert P_0)$.
\end{theorem}
The series starts at $m=k$: a pair of watermarked responses carries information only through coordinates where the underlying codewords can agree in at least $k$ places, and this information decays like $\rho^{4k}$.
 
\emph{From $\chi^2$ to $\KL$.} When $\chi^2$ is small, $\KL=\tfrac12\chi^2(1+o(1))$ holds as long as the ratio $P_{\rm WM}(x)/P_0(x)$ is close to $1$ at every $x$ that $P_0$ reaches with non-negligible probability. We do not prove this in general, but the numerical checks of Sec.~\ref{sec:val} agree with the factor $\tfrac12$ to three or four digits. Keeping only the first term $m=k$ of \eqref{eq:chi} then gives
\begin{equation}
\KL\;\approx\;\tfrac12\,\binom nk q^{-k}\rho^{4k}\;\approx\;\frac{(\vartheta\rho^4)^k}{2\,k!}.
\label{eq:KLlead}
\end{equation}
The first approximation in \eqref{eq:KLlead} drops the terms $m>k$ and is accurate when $k\vartheta\rho^4\ll1$. The second replaces $\binom nkq^{-k}$ by $\vartheta^k/k!$ and is accurate when also $k^2\ll n$. Proposition~\ref{prop:cost} uses only the first to calculate the sample cost $T$ of Algorithm \ref{alg:pair}.

The attacker receives $T$ responses and must decide whether they are watermarked or uniform. We call the ``uniform'' case the \emph{null hypothesis} (no watermark) and the ``watermarked'' case the alternative. For each of the $N=\binom T2$ pairs $(Y_i,Y_j)$ it computes the agreement count $A_{ij}$ and forms the statistic $W=\sum_{i<j}\ln\!\big(P_{\rm WM}(A_{ij})/P_0(A_{ij})\big)$. It declares ``watermarked'' if $W$ exceeds a threshold, and ``uniform'' otherwise.

\begin{proposition}[Sample cost]\label{prop:cost}
The test can make error in two ways: it can declare uniform data watermarked (false positive), or watermarked data uniform (false negative). Assume that the pair statistics are uncorrelated under the watermark. This holds exactly under the uniform null, and it is a first-order approximation under the watermark, checked in Sec.~\ref{sec:val}. Let $Q$ be the Gaussian tail function. Then, with the threshold at the midpoint of the two means of $W$, the false-positive and false-negative probabilities are both approximately $Q(z)$ once
\[
T\;\approx\;\frac{2z}{\sqrt{\KL}}.
\]
For $z=3$ (each error $\approx1.4\times10^{-3}$) this reads
\begin{equation}
T\;\approx\;\frac{6}{\sqrt{\KL}},\qquad
\log_2T\;\approx\;\tfrac12\big[1+\log_2(18/\KL)\big].
\label{eq:cost}
\end{equation}
Combined with \eqref{eq:KLlead}, this gives
$T\approx 6\sqrt2\;\rho^{-2k}\,\big(q^k/\binom nk\big)^{1/2}$.
\end{proposition}

The number of responses the attacker needs is inversely proportional to the square root of the information per pair. A small $\KL$ therefore means a secure code and a large $\KL$ a cheap attack. The factor $\rho^{-2k}$ shows that noise helps the provider exponentially in $k$.

\begin{remark}[Running time]\label{rem:time}
Proposition~\ref{prop:cost} counts responses, not computation. The running time of the test is calculated both classically and quantumly as follows.
 
\emph{Classical.} The test evaluates $A_{ij}$ on all $N=T^2/2=18/\KL$ pairs, at $O(n)$ operations per pair. The total is about $18\,n/\KL$ operations, that is, $2\log_2T+\log_2n-1$ bits. Using fewer pairs does not help. Each pair contributes a term of mean $\KL$ and variance about $2\KL$, so averaging $N'$ randomly chosen pairs has standard error $\sqrt{2\KL/N'}$. This is below the signal $\KL$ only if $N'\gtrsim 18/\KL$ (at $z=3$), which is again all the pairs. An alternative is to hash the $k$-subsets of positions, at cost about $T\binom nk$. This is cheaper than the pairwise test when $\binom nk<T$.
 
\emph{Quantum.} Suppose the $T$ responses are stored in quantum random-access memory. Quantum mean estimation~\cite{Montanaro15,BHMT02} then computes the statistic to the required accuracy with $\tilde O(1/\sqrt{\KL})=\tilde O(T)$ evaluations. The time falls to $\tilde O(nT)$, so the exponent drops to $\log_2T+\log_2n$, while the number of responses $T$ is unchanged.

\end{remark}

\begin{corollary}[Security at the decoding limit]\label{cor:nogo}
Take the extended RS code, $n=q+1$ (the case $n=q$ is identical up to the $O(1)$ term), and the largest dimension $k=k_{\rm dec}=\floor{n\rho^2}$ that a polynomial-time detector can decode (see~\eqref{eq:kdec}). With $\KL$ from \eqref{eq:KLlead}, we get
\begin{equation}
\begin{split}
\log_2 T\;=\;&k\Big[\tfrac12\log_2\tfrac ne+\log_2\tfrac1\rho\Big]\\
&+\log_2(6\sqrt2)+\tfrac14\log_2(2\pi k)+O(1+n\rho^4).
\end{split}
\label{eq:cap}
\end{equation}
\end{corollary}
The main term is $k$ times a factor of order $\log n$, so the cost grows like $2^{\Theta(n\rho^2\log n)}$. For example, at $\rho=\tfrac12$ and $n=q$, eq.~\eqref{eq:cap} gives $\log_2T\approx27$, $57$ and $126$ bits for $n=32$, $64$ and $128$. The last term $O(1+n\rho^4)$ collects the Stirling and $k^2/n$ corrections.

To reach $2^{\lambda_{\rm s}}$ security, the main term must be at least $\lambda_{\rm s}$ minus the small constants, that is, $n\rho^2\gtrsim(\lambda_{\rm s}-3)/\big[\tfrac12\log_2(n/e)+\log_2(1/\rho)\big]$. In terms of the response length, this means $B\gtrsim nL/\kappa(\tau)$ tokens. Short responses cannot reach $\lambda_{\rm s}$ bits unless $\rho$ is large, and gating raises $\rho$ but lowers $\kappa$.

The following theorem turns the cost of Proposition~\ref{prop:cost} into a function of what the provider controls: the response length $B$ (in tokens), the gate $\tau$ and the symbol size $L$. We use the quantities of Sec.~\ref{sec:channel}. Here $\kappa(\tau)$ is the number of embedded bits per token, $\bar p$ is the mean bit-flip probability of the embedded decisions, and $\eta$ is the extra substitution rate from the channel (edits or paraphrase).

\begin{theorem}[Token-budget cost]\label{thm:token}
Fix a language model and a gate $\tau$, with $\kappa$ embedded bits per token and mean bit-flip probability $\bar p$. Fix also a channel rate $\eta$, a response length of $B$ tokens and a symbol size $L$. Define
\[
\begin{aligned}
q&=2^L,\qquad n=\min\{q+1,\,B\kappa/L\},\\
\rho&=(1-\eta)(1-\bar p)^L,\qquad a=\log_2\tfrac1{1-\bar p},
\end{aligned}
\]
so that $\rho=(1-\eta)2^{-aL}$, and let $\tilde k=n\rho^2$ be the decoding limit \eqref{eq:kdec} without rounding. Assume $\tilde k\ge2$ and use the leading-term approximation \eqref{eq:KLlead}, which is accurate when $\tilde k\vartheta\rho^4\lesssim1$. Then:
\begin{enumerate}
\item[(a)] \emph{Cost,} with the real-valued $\tilde k$ in place of $k_{\rm dec}=\lfloor\tilde k\rfloor$. $\displaystyle \log_2T=\tilde k\Big[\tfrac12\log_2\tfrac qe+\log_2\tfrac1\rho\Big]+\tfrac14\log_2(2\pi\tilde k)+\log_2(6\sqrt2)+O(1+n\rho^4).$ Rounding $\tilde k$ down to $k_{\rm dec}$ lowers the right-hand side by at most one $k$-step, about $\tfrac12\log_2q+\log_2\tfrac1\rho$ bits.
\item[(b)] \emph{Closed form in $B$.} If $n=B\kappa/L\le q+1$ (all embedded bits are used), then, with $\delta=\tfrac12\log_2e-\log_2\frac1{1-\eta}$,
\begin{equation}
\begin{split}\log_2T\;\approx\;&\frac{B\kappa(1-\eta)^2}{L}\,2^{-2aL}\Big[\big(\tfrac12+a\big)L-\delta\Big]\\&+\tfrac14\log_2(2\pi\tilde k)+\log_2(6\sqrt2).\end{split}
\label{eq:token}
\end{equation}
\item[(c)] \emph{Best symbol size.} Let $L_{\min}=\min\{L:\,B\kappa/L\le2^L+1\}$ be the smallest $L$ for which all embedded bits fit, and let $\bar p<p^\ast=1-1/\sqrt2$. For $2\le L<L_{\min}$ the code length is capped at $q+1$ and the cost increases with $L$ (for this range we also assume $\bar p<0.225$, or $2^L\gg1$). For $L\ge\max\{L_{\min},\,1.5+1.1/\sqrt a\}$ the cost decreases, as $2^{-2aL}$. The best $L$ is therefore $L_{\min}-1$ or $L_{\min}$.
\item[(d)] \emph{Power law.} For $L=L_{\min}\approx\log_2(B\kappa)-\log_2\log_2(B\kappa)$ and large $B\kappa$,
\begin{equation}
\log_2T\;\approx\;\big(\tfrac12+a\big)(1-\eta)^2\,L^{2a}\,(B\kappa)^{\,1-2a}.
\label{eq:power}
\end{equation}
\end{enumerate}
\end{theorem}

\begin{remark}
Theorem~\ref{thm:chi} holds for a fixed dimension $k$. Corollary~\ref{cor:nogo} and Theorem~\ref{thm:token} apply the approximation \eqref{eq:KLlead} at $k=\tilde k=n\rho^2$, which grows with $n$. We have not proved that the resulting error is $O(1+n\rho^4)$ bits for growing $k$, so we treat these two results as accurate approximations and test them against the exact pair distribution of Theorem~\ref{thm:law} in next Sec.~\ref{sec:val}. The tests show the following.
\begin{itemize}
\item The closed form of Theorem~\ref{thm:token}(a) lies within $-2.8$ to $+6.0$ bits of the exact cost.
\item Most of this gap comes from using $\tilde k$ instead of $\lfloor\tilde k\rfloor$. One step in $k$ changes $\log_2T$ by $4$--$6$ bits.
\item The condition $\tilde k\vartheta\rho^4\lesssim1$ fails in a few cases, for example $\tilde k\vartheta\rho^4\approx3$ for Llama-2-7B at $\tau=0.4$ and $B=650$. There the closed form is still within $1.1$ bits.
\end{itemize}
\end{remark}

\begin{corollary}[No-go and token requirement]\label{cor:tokens} 
Let $p^\ast=1-1/\sqrt2\approx0.293$ and use the notation of Theorem~\ref{thm:token}.
\begin{enumerate}
\item[(i)] \emph{No-go.} If $\bar p\ge p^\ast$, then $n\rho^2\le1+2^{-L}<2$ for every $B$ and $L$. Hence no Reed--Solomon dimension $k\ge2$ is decodable by the list decoder, however long the response. (Under the exact condition $k-1<n\rho^2$, $k\ge3$ is excluded, and $k=2$ is possible only in the degenerate range $1<n\rho^2\le1+2^{-L}$.) Ungated Llama-2-7B, with measured $\bar p=0.445$, is in this regime.
\item[(ii)] \emph{Token requirement.} If $\bar p<p^\ast$, the cost grows only as $(B\cdot\kappa)^{1-2a}$, with exponent $1-2a<1$. To leading order, $2^{\lambda_{\rm s}}$ security needs
\begin{equation}
B\;\gtrsim\;\frac1\kappa\Big[\frac{\lambda_{\rm s}}{(\frac12+a)(1-\eta)^2L^{2a}}\Big]^{\frac1{1-2a}}.
\label{eq:Blambda}
\end{equation}
Thus $B$ grows polynomially in $\lambda_{\rm s}$, as $\lambda_{\rm s}^{1/(1-2a)}$, of degree $1/(1-2a)>1$ (about $2$ for $\bar p\approx0.16$).
\item[(iii)] \emph{Gate.} At fixed $B$, raising $\tau$ lowers $\bar p$ (through $a$) but also lowers $\kappa$. The best gate maximises Eq.~\eqref{eq:token} over $\tau$.
\end{enumerate}
\end{corollary}

\begin{remark}[Relation to the conjecture]
With $k=n^{1/5}$, Proposition~\ref{prop:cost} and \eqref{eq:KLlead} give $T\approx6\sqrt2\,\rho^{-2k}\sqrt{k!}=2^{\Theta(k\log k)}=2^{\Theta(n^{1/5}\log n)}$. The terms $m>k$ of the series \eqref{eq:chi}, which the leading-term approximation drops, change $\log_2T$ only by a lower-order amount (we computed $0.4$--$3$ bits for $k=10$--$160$ at $\rho=\tfrac12$, $\vartheta=1$). The cost is superpolynomial in $n$, so the attack does not contradict the conjecture, which concerns $\mathrm{poly}(n)$ samples. Our results concern the \emph{constants} at deployable sizes, not the asymptotic claim.
\end{remark}

\section{Numerical Validation}\label{sec:val}
We test the theory in three steps. We first measure the channel noise of real LLM watermarks (Sec.~\ref{sec:etaeff}). We then compare the closed forms of Sec.~\ref{sec:pair} with the exact pair law and give the attack cost at LLM operating points. Finally, we run the attack end to end on GPT-2 and Llama-2-7B watermarks (Sec.~\ref{sec:e2e}).
 
\subsection{Decoding-channel noise $\eta_{\rm eff}$ of LLM watermarks}\label{sec:etaeff}
We are going to measure the noise values used below. We cannot assume we have them in advance. By \eqref{eq:rho}, we have that
\begin{equation}
\eta_{\rm eff}\;=\;1-(1-\eta)\,\E\Big[\prod_{i=1}^{L}(1-p_i)\Big]\;\approx\;1-(1-\eta)(1-\bar p)^{L},
\label{eq:etaeff}
\end{equation}
where $p_i=\tfrac12-m_i$ is the flip probability of the $i$th embedded decision, measured from the model's next-token distribution. The mean $\bar p=\E[p_i]$ is taken over the decisions kept by the gate $m_i\ge\tau$, and $\eta\in\{0,0.1\}$ is the extra substitution rate of the channel. Even with $\eta=0$, $\eta_{\rm eff}$ is large, because an $L$-bit symbol survives only if all its bits do so. Table~\ref{tab:etaeff} lists the measured $\kappa$ and $\bar p$ on WikiText-2 and the resulting $\eta_{\rm eff}$.
 
\begin{table*}[t]
\centering
\caption{Decoding-channel noise $\eta_{\rm eff}=1-(1-\eta)(1-\bar p)^L$ for $\eta\in\{0,0.1\}$ shown with two numbers separated by $/$. These are measured on WikiText-2 in teacher mode. $\kappa$ and $\bar p$ carry 95\% bootstrap half-widths. The last two rows use the models' own samples (generation mode) at $\tau=0.2$. The dagger notation $^{\dagger}$ marks $\bar p\ge p^\ast=1-1/\sqrt2\approx0.293$, where Corollary~\ref{cor:tokens}(i) applies: no $k\ge2$ is list-decodable for any $B$ and $L$.}
\label{tab:etaeff}
\small
\begin{tabular}{@{}llccccccc@{}}
\toprule
Model & $\tau$ & $\kappa$ & $\bar p$ & $2(1-\bar p)^2$ & $L{=}4$ & $L{=}6$ & $L{=}8$ & $L{=}10$\\
\midrule
Llama-2-7B & 0 & 15.00 & 0.445$\pm$0.002$^{\dagger}$ & 0.62 & 0.905 / 0.915 & 0.971 / 0.974 & 0.991 / 0.992 & 0.997 / 0.998\\
Llama-2-7B & 0.01 & 5.28$\pm$0.16 & 0.346$\pm$0.001$^{\dagger}$ & 0.86 & 0.817 / 0.835 & 0.922 / 0.929 & 0.966 / 0.970 & 0.986 / 0.987\\
Llama-2-7B & 0.03 & 4.17$\pm$0.14 & 0.310$\pm$0.001$^{\dagger}$ & 0.95 & 0.773 / 0.796 & 0.892 / 0.903 & 0.948 / 0.954 & 0.975 / 0.978\\
Llama-2-7B & 0.05 & 3.58$\pm$0.12 & 0.285$\pm$0.001 & 1.02 & 0.738 / 0.764 & 0.866 / 0.879 & 0.931 / 0.938 & 0.965 / 0.968\\
Llama-2-7B & 0.1 & 2.66$\pm$0.09 & 0.236$\pm$0.001 & 1.17 & 0.659 / 0.693 & 0.801 / 0.821 & 0.884 / 0.895 & 0.932 / 0.939\\
Llama-2-7B & 0.2 & 1.64$\pm$0.06 & 0.161$\pm$0.001 & 1.41 & 0.505 / 0.554 & 0.651 / 0.686 & 0.755 / 0.779 & 0.827 / 0.845\\
Llama-2-7B & 0.3 & 1.00$\pm$0.04 & 0.103$\pm$0.001 & 1.61 & 0.351 / 0.416 & 0.477 / 0.530 & 0.579 / 0.621 & 0.661 / 0.695\\
Llama-2-7B & 0.4 & 0.48$\pm$0.02 & 0.051$\pm$0.001 & 1.80 & 0.188 / 0.269 & 0.268 / 0.342 & 0.341 / 0.407 & 0.406 / 0.465\\
\addlinespace
GPT-2 & 0 & 16.00 & 0.392$\pm$0.001$^{\dagger}$ & 0.74 & 0.864 / 0.877 & 0.950 / 0.955 & 0.981 / 0.983 & 0.993 / 0.994\\
GPT-2 & 0.01 & 10.17$\pm$0.06 & 0.332$\pm$0.001$^{\dagger}$ & 0.89 & 0.801 / 0.821 & 0.911 / 0.920 & 0.960 / 0.964 & 0.982 / 0.984\\
GPT-2 & 0.03 & 8.28$\pm$0.07 & 0.298$\pm$0.001$^{\dagger}$ & 0.99 & 0.757 / 0.781 & 0.880 / 0.892 & 0.941 / 0.947 & 0.971 / 0.974\\
GPT-2 & 0.05 & 7.16$\pm$0.06 & 0.273$\pm$0.001 & 1.06 & 0.720 / 0.748 & 0.852 / 0.867 & 0.922 / 0.929 & 0.958 / 0.963\\
GPT-2 & 0.1 & 5.42$\pm$0.05 & 0.223$\pm$0.000 & 1.21 & 0.635 / 0.671 & 0.779 / 0.801 & 0.867 / 0.880 & 0.919 / 0.927\\
GPT-2 & 0.2 & 3.51$\pm$0.04 & 0.151$\pm$0.000 & 1.44 & 0.480 / 0.532 & 0.625 / 0.662 & 0.729 / 0.756 & 0.805 / 0.824\\
GPT-2 & 0.3 & 2.30$\pm$0.03 & 0.097$\pm$0.000 & 1.63 & 0.336 / 0.402 & 0.459 / 0.513 & 0.559 / 0.603 & 0.641 / 0.677\\
GPT-2 & 0.4 & 1.20$\pm$0.02 & 0.049$\pm$0.000 & 1.81 & 0.183 / 0.265 & 0.262 / 0.335 & 0.332 / 0.399 & 0.397 / 0.457\\
\addlinespace
Llama-2-7B (gen.) & 0.2 & 2.17$\pm$0.17 & 0.158$\pm$0.002 & 1.42 & 0.498 / 0.549 & 0.645 / 0.680 & 0.748 / 0.774 & 0.822 / 0.840\\
GPT-2 (gen.) & 0.2 & 4.01$\pm$0.04 & 0.148$\pm$0.000 & 1.45 & 0.472 / 0.525 & 0.616 / 0.655 & 0.721 / 0.749 & 0.797 / 0.818\\
\bottomrule
\end{tabular}
\end{table*}
 
We observe that: (i) If no gating is used, the channel is unusable. At $\tau=0$, $\bar p=0.445$ (Llama-2-7B) and $0.392$ (GPT-2) both exceed $p^\ast$, and $\eta_{\rm eff}=0.91$ and $0.86$ already at $L=4$. This is the regime of Corollary~\ref{cor:tokens}(i).
(ii) There is a price to be paid for gating making the code decodable. $\bar p$ falls below $p^\ast$ at $\tau\approx0.043$ (Llama-2-7B) and $\tau\approx0.034$ (GPT-2), and larger $\tau$ lowers $\eta_{\rm eff}$. At the same time $\kappa$ drops, from $15.0$ to $0.48$ bits per token for Llama-2-7B and from $16.0$ to $1.20$ for GPT-2 ($\tau=0$ to $0.4$). This is the trade-off of Theorem~\ref{thm:token}. At $\tau=0.2$ GPT-2 has the larger $\kappa$ ($3.51$ against $1.64$) at similar $\bar p$, so it is the harder target at equal $B$.
(iii) The teacher mode used to measure $\eta_{\rm eff}$ is a close proxy for generation. The models' own samples give a larger $\kappa$ than human text ($+33\%$ Llama-2-7B, $+14\%$ GPT-2 at $\tau=0.2$) at nearly equal $\bar p$. The i.i.d.\ approximation in \eqref{eq:etaeff} changes the symbol survival by only $0.1$--$2.0\%$ at $\tau=0.2$. The costs below are therefore slightly lower than those for generated text.
 
\subsection{Exactness of the closed forms}
Table~\ref{tab:series} compares two values of the information per pair. The column ``$\KL$ exact'' is $\KL(P_{\rm WM}\Vert P_0)$ of \eqref{eq:Ddef}, computed numerically from the closed form $P_{\rm WM}$ of Theorem~\ref{thm:law}(c), including the exact finite-$q$ root-count distribution. The column ``$\KL$ series'' is $\tfrac12$ times the right-hand side of \eqref{eq:chi} with $\vartheta=n/q$. Both columns of $\log_2T$ use \eqref{eq:cost}. The cost $\log_2T$ agrees to within about $0.2$ bit on all rows shown. The series drifts only for $k$ far above the decoding limit.

\begin{table}[t]
\centering
\caption{Exact $\KL$ (Eq.~\eqref{eq:Ddef}, computed from Thm.~\ref{thm:law}) versus the series $\tfrac12\times$ the right-hand side of \eqref{eq:chi} with $\vartheta=n/q$; $\eta_{\rm eff}=1-\rho$. The last two columns apply \eqref{eq:cost} to the exact $\KL$ (``ex.'') and to the series $\KL$ (``ser.'').}
\label{tab:series}
\footnotesize\setlength{\tabcolsep}{3.2pt}
\begin{tabular}{@{}rrrrrrrr@{}}
\toprule
$n$ & $q$ & $k$ & $\eta_{\rm eff}$ & $\KL$ exact & $\KL$ series & $\log_2T$ ex. & ser.\\
\midrule
257 & 256 & 3 & 0.816 & 1.30e-10 & 1.27e-10 & 19.01 & 19.02\\
257 & 256 & 8 & 0.816 & 3.74e-29 & 3.84e-29 & 49.80 & 49.78\\
513 & 512 & 17 & 0.816 & 1.18e-65 & 1.51e-65 & 110.43 & 110.25\\
257 & 256 & 8 & 0.500 & 3.90e-15 & 4.49e-15 & 26.52 & 26.42\\
32 & 32 & 3 & 0.550 & 5.64e-06 & 6.29e-06 & 11.30 & 11.22\\
1024 & 1024 & 16 & 0.700 & 8.10e-48 & 9.24e-48 & 80.80 & 80.71\\
\bottomrule
\end{tabular}
\end{table}
 
\subsection{Security at the decoding limit}
Table~\ref{tab:cap} compares \eqref{eq:cap} with the exact cost at $k=k_{\rm dec}=\floor{n\rho^2}$. The main term $k\big[\tfrac12\log_2(n/e)+\log_2(1/\rho)\big]$ alone is $4$--$8$ bits low. Adding the lower-order terms $\log_2(6\sqrt2)+\tfrac14\log_2(2\pi k)$ brings it within $0.5$--$2.5$ bits of the exact value, and the remaining gap is the $O(1+n\rho^4)$ term of \eqref{eq:cap}.

\begin{table}[!t]
\centering
\caption{Cost at the decoding limit $k=\floor{n\rho^2}$ (extended RS, $n=q+1$). ``Main term'' is $k[\frac12\log_2(n/e)+\log_2(1/\rho)]$, the first line of \eqref{eq:cap}. ``+ lower-order'' adds $\log_2(6\sqrt2)+\frac14\log_2(2\pi k)$, so it is the right-hand side of \eqref{eq:cap} without the $O(1+n\rho^4)$ term.}
\label{tab:cap}
\footnotesize\setlength{\tabcolsep}{3.2pt}
\begin{tabular}{@{}rrrrrrr@{}}
\toprule
$n$ & $\rho$ & $k_{\rm dec}$ & exact $\log_2T$ & main term & + lower-order\\
\midrule
257 & 0.184 & 8 & 49.8 & 45.8 & 50.3\\
257 & 0.300 & 23 & 120.8 & 115.4 & 120.3\\
257 & 0.400 & 41 & 195.6 & 188.7 & 193.8\\
513 & 0.250 & 32 & 190.5 & 185.0 & 190.0\\
1025 & 0.300 & 92 & 561.3 & 553.5 & 558.9\\
\bottomrule
\end{tabular}
\end{table}

\subsection{LLM operating points}
\begin{figure}[t]
\centering
\includegraphics[width=\columnwidth]{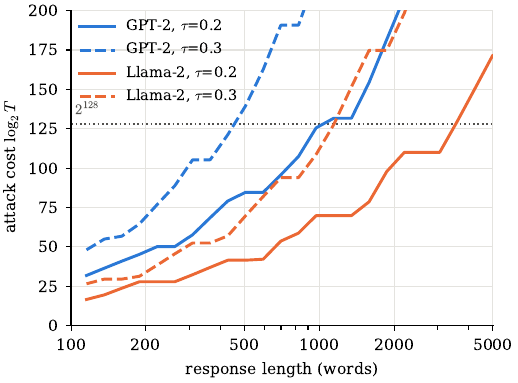}
\caption{Attack cost $\log_2T$ against response length, from the exact pair distribution at $k=k_{\rm dec}$ with the best $L$ at each length ($\eta=0.1$, teacher-mode $\kappa,\bar p$ of Table~\ref{tab:etaeff}; $1$ word $\approx1.3$ tokens). The dotted line is $2^{128}$, and the steps are the jumps of $k_{\rm dec}$. At $\tau=0.2$ about $1000$ words (GPT-2) and $3500$ words (Llama-2-7B) are needed, and a higher gate $\tau=0.3$ needs fewer.}
\label{fig:cost}
\end{figure}

Table~\ref{tab:llm} applies the exact pair distribution at $k=k_{\rm dec}$ to the teacher-mode statistics of Table~\ref{tab:etaeff}, so both models are compared under the same protocol. For each token budget $B$ we choose the $L$ that maximises the cost, subject to $n=\min(q+1,\floor{B\kappa/L})$ and $\rho=(1-\eta)(1-\bar p)^L$. The budgets $400$ and $650$ tokens are about $300$ and $500$ words.
We observe that short responses are not secure against this test, (see Fig.~\ref{fig:cost}). At $\tau=0.2$ a 300-word response gives a cost of only $2^{29}$ to $2^{58}$ and a 500-word response $2^{39}$ to $2^{85}$, far below $2^{128}$. Reaching $2^{128}$ at this gate needs about 1000--1600 words for GPT-2 and 3500--4400 for Llama-2-7B; a higher gate shortens this to roughly 400--560 (GPT-2) and 1010--1430 (Llama-2-7B) words at $\tau\ge0.3$, where $\kappa$ is smaller but $\bar p$ is much lower.
 
\begin{table*}[t]
\centering
\caption{$\log_2T$ (attack cost, number of responses) computed from the exact pair distribution at the decoding limit, and the response length needed for $2^{128}$. Values in each cell are for $\eta=0.1\,/\,0.2$.}
\label{tab:llm}
\small
\begin{tabular}{@{}llccccc@{}}
\toprule
Model & $\tau$ & $\kappa$ & $\bar p$ & $B{=}400$ tok ($\approx$300 words) & $B{=}650$ tok ($\approx$500 words) & $B_{\min}$ for $2^{128}$ (tokens; words)\\
\midrule
GPT-2    & 0.2 & 3.51 & 0.151 & $2^{58}$ / $2^{48}$ & $2^{85}$ / $2^{64}$ & 1267 / 2117; 976 / 1630\\
GPT-2    & 0.3 & 2.30 & 0.097 & $2^{105}$ / $2^{84}$ & $2^{140}$ / $2^{115}$ & 595 / 724; 458 / 557\\
GPT-2    & 0.4 & 1.20 & 0.049 & $2^{101}$ / $2^{82}$ & $2^{162}$ / $2^{134}$ & 514 / 630; 396 / 485\\
\addlinespace
Llama-2-7B  & 0.2 & 1.64 & 0.161 & $2^{32}$ / $2^{29}$ & $2^{42}$ / $2^{39}$ & 4537 / 5744; 3493 / 4423\\
Llama-2-7B  & 0.3 & 1.00 & 0.103 & $2^{53}$ / $2^{44}$ & $2^{70}$ / $2^{55}$ & 1520 / 1856; 1170 / 1429\\
Llama-2-7B  & 0.4 & 0.48 & 0.051 & $2^{43}$ / $2^{35}$ & $2^{71}$ / $2^{58}$ & 1313 / 1619; 1011 / 1247\\
\bottomrule
\end{tabular}
\end{table*}
 
\subsection{Token-budget distribution}
Table~\ref{tab:token} compares the closed form of Theorem~\ref{thm:token}(a), at the best $L\in\{L_{\min}-1,L_{\min}\}$, with the cost from the exact pair distribution, for the same inputs as Table~\ref{tab:llm}. The closed form lies within $-2.8$ to $+6.0$ bits of the exact cost in all 24 cells (closed form minus exact). The exact cost is a staircase in the inputs, jumping by about $6$ bits when $n\rho^2$ crosses an integer, which explains the largest gaps. The power law \eqref{eq:power} reproduces the growth exponent $1-2a$, but its level is $7$--$31\%$ too high, so we use it as a scaling guide. Tables~\ref{tab:llm} and~\ref{tab:token} are computed from the printed values of Table~\ref{tab:etaeff}.

\begin{table}[!t]
\centering
\caption{$\log_2T$ from the exact pair distribution vs.\ the closed form of Theorem~\ref{thm:token}(a); each cell reads exact / closed form, at the best $L$; inputs as in Table~\ref{tab:llm}.}
\label{tab:token}
\footnotesize\setlength{\tabcolsep}{3.2pt}
\begin{tabular}{@{}llrccc@{}}
\toprule
Model & $\tau$ & $B$ & $1-2a$ & $\eta{=}0.1$ & $\eta{=}0.2$\\
\midrule
GPT-2 & 0.2 & 400  & 0.53 & 57.7 / 59.7 & 48.4 / 49.4\\
  & & 650  &      & 84.7 / 85.4 & 64.4 / 70.4\\
GPT-2 & 0.3 & 400  & 0.71 & 105.3 / 104.1 & 83.6 / 86.6\\
  & & 650  &      & 139.7 / 141.3 & 115.5 / 116.6\\
GPT-2 & 0.4 & 400  & 0.86 & 101.2 / 99.5 & 82.3 / 83.2\\
  & & 650  &      & 161.5 / 158.7 & 134.5 / 132.4\\
\addlinespace
Llama-2-7B & 0.2 & 400  & 0.49 & 32.4 / 35.0 & 28.6 / 29.4\\
  & & 650  &      & 41.6 / 46.6 & 38.6 / 38.9\\
Llama-2-7B & 0.3 & 400  & 0.69 & 52.5 / 52.9 & 43.9 / 44.6\\
  & & 650  &      & 69.9 / 70.9 & 54.9 / 59.1\\
Llama-2-7B & 0.4 & 400  & 0.85 & 43.3 / 44.6 & 35.4 / 38.0\\
  & & 650  &      & 70.8 / 69.7 & 57.7 / 59.0\\
\bottomrule
\end{tabular}
\end{table}

\subsection{End-to-end test on GPT-2 and Llama-2-7B}\label{sec:e2e}
\begin{table*}[t]
\centering
\caption{End-to-end attack on GPT-2 and Llama-2-7B with the extended RS code ($q=16$, $n=17$, $L=4$, $\tau=0.2$). ``Predicted'' uses $T\approx6/\sqrt{\KL}$ with the exact distribution of Theorem~\ref{thm:law}, evaluated at the measured $\eta_{\rm eff}$. ``Distinct codewords'' uses $P_{\rm WM}$ of Theorem~\ref{thm:law}(c), which ignores pairs of responses that carry the same codeword. ``All pairs'' uses the full distribution of $A$ in Theorem~\ref{thm:law}(c), which includes those pairs (probability $q^{-k}$) and so gives a slightly lower cost. ``Empirical'' extrapolates the set-level AUC to the $z=4.24$ criterion ($\pm0.3$ bit). Attack AUC is the keyless pair test on sets of $T$ responses, computed on a held-out half; $^{\ast}$ marks AUC $1.00$ inferred from smaller $T$. Key-holder AUC is the statistic $M$ on one response with the key; the two AUCs are not comparable. For $k=4$, entries are means of four seeds. $k>k_{\rm dec}$ is beyond list decoding.}
\label{tab:e2e}
\footnotesize
\setlength{\tabcolsep}{4pt}
\begin{tabular}{@{}l ccc ccc cc@{}}
\toprule
 & \multicolumn{6}{c}{GPT-2} & \multicolumn{2}{c}{Llama-2-7B}\\
\cmidrule(lr){2-7}\cmidrule(l){8-9}
 & \multicolumn{3}{c}{$\eta=0$} & \multicolumn{3}{c}{$\eta=0.1$} & \multicolumn{2}{c}{$\eta=0$}\\
\cmidrule(lr){2-4}\cmidrule(lr){5-7}\cmidrule(l){8-9}
$L,\,k$ & 4, 2 & 4, 3 & 4, 4 & 4, 2 & 4, 3 & 4, 4 & 4, 2 & 4, 3\\
\midrule
$k_{\rm dec}=\lfloor n\rho^2\rfloor$ & 4 & 4 & 4 & 3 & 3 & 3 & 4 & 4\\
$\eta_{\rm eff}$ (measured) & 0.472 & 0.471 & 0.472 & 0.519 & 0.518 & 0.520 & 0.497 & 0.500\\
predicted $\KL$ (distinct codewords) & $1.8{\times}10^{-3}$ & $4.7{\times}10^{-5}$ & $8.0{\times}10^{-7}$ & $8.6{\times}10^{-4}$ & $1.6{\times}10^{-5}$ & $1.9{\times}10^{-7}$ & $1.2{\times}10^{-3}$ & $2.4{\times}10^{-5}$\\
predicted $\log_2T$ (distinct codewords) & 7.1 & 9.8 & 12.7 & 7.7 & 10.6 & 13.8 & 7.4 & 10.3\\
predicted $\log_2T$ (all pairs) & 7.1 & 9.6 & 12.4 & 7.5 & 10.3 & 13.4 & 7.3 & 10.0\\
empirical $\log_2T$ & 6.9 & 9.7 & 13.0 & 7.6 & 10.5 & 13.2 & 7.3 & 10.1\\
attack AUC (no key), $T{=}1000$ & $1.00^{\ast}$ & 1.00 & 0.65 & $1.00^{\ast}$ & 1.00 & 0.66 & -- & --\\
attack AUC (no key), $T{=}3000$ & -- & -- & 0.97 & -- & -- & 0.93 & -- & --\\
attack AUC (no key), $T{=}250$ & -- & -- & -- & -- & -- & -- & 1.00 & 0.77\\
attack AUC (no key), $T{=}500$ & -- & -- & -- & -- & -- & -- & 1.00 & 0.98\\
key-holder AUC (key, 1 resp.) & 0.98 & 0.93 & 0.86 & 0.95 & 0.88 & 0.79 & 0.97 & 0.91\\
\bottomrule
\end{tabular}
\end{table*}
We embedded a random permuted-RS codeword in each GPT-2 and Llama-2-7B response (optimal coupling, gate $\tau=0.2$) and attacked the extracted symbols with the pair test. The null (no-watermark) reference is real unwatermarked GPT-2 text, and the attacker is oracle-aligned (it knows which decisions carry bits). The code is the extended RS code with $n=q+1=17$ symbols ($L=4$). We ran $k=2,3,4$ on GPT-2, at $\eta=0$ and at extra channel noise $\eta=0.1$, and $k=2,3$ on Llama-2-7B at $\eta=0$.

\emph{Metrics.} Each detector outputs a statistic $S$ (larger means ``more watermarked''). Let $S_{\rm wm}$ be the value of $S$ on a watermarked input, and $S_{\rm ctrl}$ its value on an independent control input drawn from real unwatermarked text. An input is a set of $T$ responses for the pair test and a single response for the key-holder detector. We report the area under the Receiver Operating Characteristic (ROC) curve,
\begin{equation}\label{eq:auc}
\mathrm{AUC}=\Pr[S_{\rm wm}>S_{\rm ctrl}]+\tfrac12\Pr[S_{\rm wm}=S_{\rm ctrl}],
\end{equation}
so that $0\le\mathrm{AUC}\le1$. It is $\tfrac12$ when the statistic cannot tell watermarked from control inputs (random guessing) and $1$ for perfect separation. A larger AUC therefore means a more effective attack (or detector), and for the attack it means a less secure code. Values below $\tfrac12$ would only mean that the statistic has the wrong sign, and do not occur here. The keyless \emph{pair-test attack} is Algorithm~\ref{alg:pair} applied to a set of $T$ responses, with two differences. First, because the control is real text, the table $h$ is estimated from the first half of the data (histograms of $A$ for watermarked and for control pairs, with additive smoothing) instead of from Theorem~\ref{thm:law}(c). Second, instead of thresholding $W$, we report the AUC of $W$ between watermarked and control sets drawn from the held-out half. The theoretical $P_{\rm WM}$ enters only the ``predicted'' rows. We estimate $h$ from data because real unwatermarked text has biased, correlated symbols, so its agreement counts need not be Binomial$(n,1/q)$. We report the AUC because error rates near $10^{-3}$ cannot be measured directly from a few thousand responses, whereas the AUC curve in $T$ can be extrapolated to the $z=4.24$ criterion. The AUC is thus an evaluation measure that needs labelled watermarked and control sets; a deployed attacker would use the threshold of Algorithm~\ref{alg:pair}.

The \emph{key-holder detector} knows $(\sigma,\pi_1,\dots,\pi_n)$ and judges one response by $M$, its maximum number of agreements with any codeword after undoing the key. The two AUCs answer different questions and should not be compared as scores. The key-holder AUC measures how well the provider, who knows the key, detects the watermark in one response. The keyless AUC measures how well an attacker without the key distinguishes watermarked from uniform responses, and it depends on the number $T$ of responses collected, so it is $\tfrac12$ for small $T$ and tends to $1$ as $T$ grows. The key-holder rows are therefore a reference that the watermark remains detectable with the key, not a baseline for the attack. The keyless AUC is meaningful only together with $T$, and its $T$ is the cost of Proposition~\ref{prop:cost}. To compare with Proposition~\ref{prop:cost}, we extrapolate the measured AUC (which grows with $T$) to $z=4.24=3\sqrt2$. This is the point that corresponds to the $z=3$ operating point of the proposition, because the AUC compares two independent statistics.

\emph{False alarms of $M$.} For a uniform word and $m\ge k$, a fixed set of $m$ positions is matched by some codeword with probability $q^{k-m}$, so
\begin{equation}\label{eq:mnull}
\Pr[M\ge m]\;\le\;\tbinom{n}{m}\,q^{k-m}\;\le\;2^{n}\,q^{k-m}.
\end{equation}
This is the counting bound behind the proximity-gap analysis of Crites and Stewart~\cite{CS25a} (their Lemma~1), and the simulated false-alarm rates agree with it within a factor of about two. In the PRSC regime ($n\approx q$, $k=\lambda_{\rm s}^{1/5}$) the bound is $q^{-\Omega(n)}$ once $m-k=\Omega(n)$, so the key-holder test is not the bottleneck; the keyless attack is.
 
\emph{Results} (Table~\ref{tab:e2e}). The channel model holds in every run (bit error within $0.0005$ and symbol survival within $1\%$ of the prediction). The empirical control distribution differs slightly from the Binomial$(n,1/q)$ null, so we use the empirical control as the null. The measured cost agrees with the predicted cost within $0.3$ bit in nearly all columns, including both Llama-2-7B columns. The attack separates watermarked from control responses with $T=1000$ responses for GPT-2 ($k=2,3$; AUC $1.00$) and $T=500$ for Llama-2-7B (AUC $\ge0.98$).
The one wider column is GPT-2 with $k=4$ and $\eta=0.1$. Over four seeds the estimate ranges over $12.2$--$14.2$ (mean $13.2$), against $13.8$ predicted, because the $q^k=65,536$ codewords make the AUC curve lumpy. Extra channel noise raises the cost as predicted ($+0.6$ to $+1.1$ bits predicted; $+0.7$ and $+0.8$ bits measured at $k=2,3$). Identical codewords lower the predicted cost by at most $0.4$ bit.

\section{Conclusion}
We gave a keyless pair-agreement test for the permuted Reed--Solomon code (PRSC) with the exact distribution of the agreement count, the information per pair, and the sample cost $T\approx6/\sqrt{\KL}$. The cost $2^{\Theta(n\rho^2\log n)}$ at the decoding limit is superpolynomial in $n$, so the test does not contradict the PRSC conjecture. In tokens, once $\bar p\ge1-1/\sqrt2$ no dimension $k\ge2$ is decodable, and at gate $\tau=0.2$ responses of 300--500 words cost only $2^{29}$ to $2^{85}$ responses to attack, while $2^{128}$ needs about 1000--4400 words. End-to-end tests agree with the predicted cost within about $0.3$ bit in most settings. Future work includes an empirical null on real text, soft-decision and list-recovery detectors, and the edit-robust variants.

{\footnotesize
\bibliographystyle{IEEEtran}
\bibliography{IEEEabrv,LIUBIB}}

\appendix
\section{Proofs}
We prove Theorems~\ref{thm:law}--\ref{thm:token} in order. The corollaries follow by substitution.

\begin{IEEEproof}[Proof of Theorem~\ref{thm:law}]
\emph{(a)} The polynomials $f,f'$ are independent and uniform on the $q^k$-element space $\F_q[x]_{<k}$, so $g=f-f'$ is uniform on it and $\Pr[g=0]=q^{-k}$.
Take a set $\Gamma\subseteq\mathcal E$ of $m\le k$ evaluation points. By the MDS property \cite{MS77}, the evaluation map $g\mapsto(g(\alpha))_{\alpha\in\Gamma}$ is onto $\F_q^m$, so exactly $q^{k-m}$ polynomials vanish on $\Gamma$, and $q^{k-m}-1$ of them are nonzero. For $m\ge k$ only $g=0$ vanishes on $\Gamma$, which gives $R\le k-1$ when $g\ne0$. Dividing by $q^k-1$ shows that $\Gamma$ consists of roots with probability $\psi_m$. The number $\binom Rm$ counts the $m$-subsets of $\mathcal E$ on which $g$ vanishes. Summing over the $\binom nm$ subsets $\Gamma$ gives $\E[\binom Rm\mid g\ne0]=\binom nm\psi_m\mathbf 1[m<k]$. Applying the Binomial inversion \cite{Feller68}, $\Pr[R=r]=\sum_{m\ge r}(-1)^{m-r}\binom mr\E[\binom Rm]$, it gives the formula for $\Pr[R=r\mid g\ne0]$.

\emph{(b)} Condition on $(f,f',K)$, which fixes the difference polynomial $g=f-f'$, the set of roots of $g$ is $\mathcal R=\{\alpha\in\mathcal E: g(\alpha)=0\}\subseteq\mathcal E$ (the evaluation points where $g$ vanishes) with $|\mathcal R|=R$, and the key $K=(\sigma,\pi_1,\dots,\pi_n)$. We identify coordinates with their evaluation points. Position $j$ of the two responses carries the symbols $\pi_j(f(\alpha))$ and $\pi_j(f'(\alpha))$ with $\alpha=\sigma(j)$, before noise. Since $\pi_j$ is a bijection, these two symbols are equal exactly when $g(\alpha)=0$, that is, when $\sigma(j)\in\mathcal R$. Let $I_j$ be the indicator that $Y_j=Y'_j$, so that $A=\sum_jI_j$.

At position $j$ each of the two symbols is kept independently with probability $\rho$, so both are kept with probability $s=\rho^2$.
\begin{itemize}
\item If both are kept, then $I_j=1$ when $\sigma(j)\in\mathcal R$ and $I_j=0$ otherwise.
\item If at least one is replaced, that symbol is uniform on $\F_q$ and independent of the other symbol, so $\Pr[I_j=1]=1/q$ exactly.
\end{itemize}

Combining the two cases by the law of total probability (both kept with probability $s$, otherwise at least one replaced with probability $1-s$), and writing $\gamma_1,\gamma_0$ for the definitions in (b), we have that
\[
\Pr[I_j=1]=\begin{cases}s\cdot1+(1-s)\tfrac1q=\gamma_1, & \sigma(j)\in\mathcal R,\\[2pt] s\cdot0+(1-s)\tfrac1q=\gamma_0, & \sigma(j)\notin\mathcal R.\end{cases}
\]

\emph{Putting the indicators together.} Given $(f,f',K)$, the survival events and the replacement symbols at different positions are independent, so $I_1,\dots,I_n$ are independent. Exactly $R$ positions have $\sigma(j)\in\mathcal R$, because $\sigma$ is a bijection. Hence $A=\sum_jI_j$ is the sum of $R$ independent Bernoulli$(\gamma_{\rm r})$ variables and $n-R$ independent Bernoulli$(\gamma_{\rm n})$ variables. This distribution depends on $(f,f',K)$ only through $R$. Conditioning on $R=r$ gives the two independent binomials in (b), and $g=0$ is the case $r=n$.
 
\emph{(c)} We now average the conditional distribution in (b) over $R$, using the weights in (a). The identical-codeword event $g=0$ has probability $q^{-k}$ and gives $\Bin(n,\gamma_{\rm r})$. Given $g\ne0$, $R=r$ has probability $\Pr[R=r\mid g\ne0]$, and $A$ is then the sum of the two independent binomials in (b), whose distribution is the convolution in the formula.
 
\emph{(d)} By (b), the conditional distribution of $A$ given $(f,f',K)$ depends only on $R$, the number of roots of $g$ in $\mathcal E$. This number does not depend on $\sigma$, which only permutes the positions, and it does not depend on the $\pi_j$. Hence the distribution of $A$ is the same for every key $K$. For two uniform words, the $n$ agreement indicators are i.i.d.\ Bernoulli$(1/q)$, so $A\sim\Bin(n,1/q)$.
\end{IEEEproof}

\begin{IEEEproof}[Proof of Theorem~\ref{thm:chi}]
We first compute the moments of $A$. We then expand $\chi^2$ in Charlier polynomials (the orthogonal polynomials of the Poisson distribution \cite{KLS10}), and sum the resulting series to get to a closed form.

\emph{Step 0: the limit.} By Theorem~\ref{thm:law}(b),(c), $P_{\rm WM}$ is the mixture over $R$ (with $R\le k-1$ by (a)) of the distribution of $A=A_1+A_2$, where $A_1\sim\Bin(R,\gamma_1)$ and $A_2\sim\Bin(n-R,\gamma_0)$ are independent. We only need the structure of this mixture here, not its explicit double sum. Since $\gamma_0\le1/q$, pointwise, $\Pr[A_2=y]\le\binom ny q^{-y}\le C_\vartheta P_0(y)$, so $P_{\rm WM}(x)\le C_\vartheta\sum_{j<k}P_0(x-j)$. For $x\le n/2$ we have $P_0(x-j)/P_0(x)\le(Cx)^{j}$. For $x>n/2$ the ratio is at most polynomial in $n$, while $P_0(x-j)$ is smaller than any fixed power of $1/n$, uniformly in $q$. Hence $P_{\rm WM}/P_0$ grows at most polynomially in $x$, and the tails of $\sum_xP_{\rm WM}^2/P_0$ are uniformly summable. As $q\to\infty$, $\gamma_1\to s$ and $n\gamma_0\to(1-s)\vartheta$, so $A_1\to\Bin(R,s)$ and $A_2\to\Poi((1-s)\vartheta)$. Let $x^{\underline u}=x(x-1)\cdots(x-u+1)$ be the falling factorial. We have that $R$ has finite support $\{0,\dots,k-1\}$ and, by Theorem~\ref{thm:law}(a), its factorial moments 
\begin{equation}\label{eq:fac}
\E[R^{\underline u}]=n^{\underline u}\psi_u
\end{equation}
tend to $\vartheta^u$ for $u<k$, so its distribution converges. The \eqref{eq:fac} holds because $R^{\underline u}$ counts the ordered $u$-tuples of distinct evaluation points that are all roots of $g$: there are $n^{\underline u}$ such tuples, and each is all roots with probability $\psi_u$. A distribution on $\{0,\dots,k-1\}$ is determined by its first $k-1$ factorial moments, which gives the convergence. Hence, $P_{\rm WM}$ converges pointwise to a limit distribution $P_\infty$ with the same domination. By dominated convergence, $\chi^2(P_{\rm WM}\Vert P_0)\to\chi^2(P_\infty\Vert\Poi(\vartheta))$, and the moments of $P_\infty$ are the limits of those in Step 1. We compute this limit next.

\emph{Step 1:} For a set $U$ of $j$ positions, Theorem~\ref{thm:law}(a),(b) give
\[
\Pr[\text{all of }U\text{ agree}\mid g\ne0]=\sum_{u=0}^{j}\binom ju s^u(1-s)^{j-u}q^{-(j-u)}\psi_u,
\]
where $u$ is the number of positions in which both symbols survive and $\psi_u$ is the probability that $g$ vanishes at $u$ given points. For $u<k$ we have $\psi_u=q^{-u}(1+O(1/q))$, and for $u\ge k$ we have $\psi_u=0$. Let $\mu_j=\E_{\rm WM}[A^{\underline j}]$ and $\nu_j=\E_0[A^{\underline j}]=n^{\underline j}q^{-j}\to\vartheta^j$. Then
\begin{equation}
\mu_j=\nu_j\,(1-b_j+O(1/q)),\qquad b_j=\Pr[\Bin(j,s)\ge k].
\end{equation}
 
\emph{Step 2:} The Charlier polynomials $c_m(x)=\sum_{j=0}^m(-1)^{m-j}\binom mj x^{\underline j}\vartheta^{-j}$ are orthogonal and complete in $L^2(\Poi(\vartheta))$, with $\|c_m\|^2=m!/\vartheta^m$. Hence
$\chi^2(P_{\rm WM}\Vert\Poi(\vartheta))=\sum_{m\ge1}\E_{\rm WM}[c_m]^2\vartheta^m/m!$.
By Step 1 and $\sum_j(-1)^{m-j}\binom mj=0$ for $m\ge1$,
\begin{equation}
\E_{\rm WM}[c_m]=-\sum_{j}(-1)^{m-j}\binom mj b_j .
\end{equation}
 
\emph{Step 3:} Expand $b_j=\sum_{u\ge k}\binom ju s^u(1-s)^{j-u}$ and use $\binom mj\binom ju=\binom mu\binom{m-u}{j-u}$:
\[
\sum_{j}(-1)^{m-j}\binom mj\binom ju (1-s)^{j-u}=\binom mu(-s)^{m-u}.
\]
Therefore
$\E_{\rm WM}[c_m]=-s^m\sum_{u=k}^m(-1)^{m-u}\binom mu=\pm s^m\binom{m-1}{k-1}$,
using $\sum_{u=k}^m(-1)^{u-k}\binom mu=\binom{m-1}{k-1}$. This vanishes for $m<k$, and with $s=\rho^2$ it gives \eqref{eq:chi}.
 
\emph{Step 4:} The inequality $\KL\le\ln(1+\chi^2)\le\chi^2$ is standard \cite[Lemma~2.7]{Tsybakov09}. For \eqref{eq:KLlead}, note that the $m=k$ term of \eqref{eq:chi} is $(\vartheta\rho^4)^k/k!$ and that successive terms have ratio about $k\vartheta\rho^4$, so the first term dominates when $k\vartheta\rho^4\ll1$. The finite-$n$ form $\tfrac12\binom nkq^{-k}\rho^{4k}$ in \eqref{eq:KLlead} is the same first term, since $\binom nkq^{-k}\to\vartheta^k/k!$ for fixed $k$; we use it for growing $k$ as an approximation. The factor $\tfrac12$ between $\KL$ and $\chi^2$ is the numerical observation discussed after Theorem~\ref{thm:chi}.
\end{IEEEproof}

\begin{IEEEproof}[Proof sketch of Proposition~\ref{prop:cost}]
When the responses are uniform (the null hypothesis), the pair counts are pairwise independent, because $A_{ij}\sim\Bin(n,1/q)$ given $Y_i$, so $\mathrm{Var}_0W=N\,\mathrm{Var}_0h$. 
In the weak-signal regime, $\E_{\rm WM}h\approx-\E_0h\approx\KL$ and $\mathrm{Var}\,h\approx2\KL$. Under the watermark, two pairs that share a response are correlated only through the noise pattern of that response. This is a second-order effect in the signal. We neglect it, and the experiments of Sec.~\ref{sec:val} support this. With the threshold at the midpoint, both errors are about $Q\big(\sqrt{N\KL/2}\big)$. Setting $N=T(T-1)/2\approx T^2/2$ gives $T\approx2z/\sqrt{\KL}$.
\end{IEEEproof}
 
\begin{IEEEproof}[Proof of Corollary~\ref{cor:nogo}]
Let $n\in\{q,q+1\}$. Then $\binom nkq^{-k}=\frac1{k!}\,(n/q)^k\prod_{i<k}(1-\tfrac in)$, so Proposition~\ref{prop:cost} and \eqref{eq:KLlead} give
\[
\log_2T=\log_2(6\sqrt2)+\tfrac12\log_2k!+2k\log_2(1/\rho)+\varepsilon,
\]
where $\varepsilon=-\tfrac k2\log_2(n/q)-\tfrac12\sum_{i<k}\log_2(1-\tfrac in)$. Since $n/q\le1+1/q$, the first part of $\varepsilon$ is $O(k/q)=O(\rho^2)$ bits. The second part is $O(k^2/n)=O(n\rho^4)$ bits. Both lie inside the $O(1+n\rho^4)$ term.
Stirling's formula gives $\tfrac12\log_2k!=\tfrac k2\log_2(k/e)+\tfrac14\log_2(2\pi k)+o(1)$. Let $\tilde k=n\rho^2$, so that $k=\lfloor\tilde k\rfloor$ and $k\le\tilde k<k+1$. Since $2\log_2(1/\rho)=\log_2(n/\tilde k)$, we have
\[
\tfrac k2\log_2\tfrac ke+2k\log_2\tfrac1\rho=k\Big[\tfrac12\log_2\tfrac ne+\log_2\tfrac1\rho\Big]-\tfrac k2\log_2\tfrac{\tilde k}{k},
\]
and $0\le\tfrac k2\log_2\tfrac{\tilde k}{k}\le\tfrac k2\log_2(1+\tfrac1k)<\tfrac1{2\ln2}<0.73$ bit. This gives \eqref{eq:cap}.
The condition on $n\rho^2$ in the text after the corollary follows by solving for $k$ and dropping the nonnegative $\tfrac14\log_2$ term, and $B=nL/\kappa$ comes from Sec.~\ref{sec:channel}.
\end{IEEEproof}
 
\begin{IEEEproof}[Proof of Theorem~\ref{thm:token}]
\emph{(a)} This follows the proof of Corollary~\ref{cor:nogo}, with $\tilde k=n\rho^2$ in place of $\lfloor n\rho^2\rfloor$ and $q$ in place of $n$ inside the logarithm. By Proposition~\ref{prop:cost} and \eqref{eq:KLlead}, $\log_2T=\log_2(6\sqrt2)+\tfrac12\log_2\!\big(q^k/\binom nk\big)+2k\log_2\frac1\rho$.
Now $q^k/\binom nk=k!\,(q/n)^k\,e^{O(k^2/n)}$, and Stirling's formula gives $\tfrac12\log_2k!=\tfrac k2\log_2\frac ke+\tfrac14\log_2(2\pi k)+o(1)$.
With $k=\tilde k=n\rho^2$ (reading $k!$ as $\Gamma(k+1)$ for real $k$),
\[
\begin{aligned}
\tfrac k2\log_2\!\Big(\tfrac{k}{e}\tfrac qn\Big)+2k\log_2\tfrac1\rho&=\tfrac k2\log_2\tfrac{\rho^2q}{e}+2k\log_2\tfrac1\rho\\
&=k\Big[\tfrac12\log_2\tfrac qe+\log_2\tfrac1\rho\Big].
\end{aligned}
\]
The correction $O(k^2/n)=O(n\rho^4)$ is the error term in the statement.

\emph{(b)} Insert $\tilde k=(B\kappa/L)(1-\eta)^2(1-\bar p)^{2L}=(B\kappa/L)(1-\eta)^22^{-2aL}$, $\log_2\frac1\rho=\log_2\frac1{1-\eta}+aL$ and $\tfrac12\log_2\frac qe=\tfrac12(L-\log_2e)$ into (a). The bracket becomes $(\tfrac12+a)L-\delta$.
 
\emph{(c)} We treat the two ranges of $L$ separately.
 
\emph{Small $L$ ($L<L_{\min}$).} Here $n=2^L+1$, so $\tilde k=(2^L+1)(1-\eta)^2(1-\bar p)^{2L}$ grows with $L$ by the factor $\frac{2^{L+1}+1}{2^L+1}(1-\bar p)^2$ per unit step. This factor tends to $2(1-\bar p)^2$, which exceeds $1$ exactly when $\bar p<p^\ast$. Since $\frac{2^{L+1}+1}{2^L+1}\ge\frac53$, it exceeds $1$ for every $L\ge1$ whenever $\bar p<1-\sqrt{3/5}\approx0.225$ (all rows of Table~\ref{tab:llm}). The bracket in (a) grows linearly in $L$ and is positive for $L\ge2$ (where $q\ge4>e$ and $\rho<1$ give $\tfrac12\log_2(q/e)>0$). So the cost increases for $2\le L<L_{\min}$.
 
\emph{Large $L$ ($L\ge L_{\min}$).} Here $n=B\kappa/L$, and by (b) the cost is $\Lambda\,\Gamma(L)+\tfrac14\log_2(2\pi\tilde k)+\mathrm{const}$, with $\Lambda=B\kappa(1-\eta)^2$ and $\Gamma(L)=2^{-2aL}\big[(\tfrac12+a)-\delta/L\big]$. The term $\tilde k=\Lambda\,2^{-2aL}/L$ decreases in $L$, so $\tfrac14\log_2(2\pi\tilde k)$ decreases as well. For $\Gamma$,
\[
\frac{d}{dL}\ln \Gamma=-2a\ln2+\frac{\delta}{L^2[(\frac12+a)-\delta/L]}.
\]
If $\delta\le0$ (that is, $\eta\ge1-2^{-\frac12\log_2e}\approx0.39$), this is negative for all $L>0$. If $0<\delta\le\tfrac12\log_2e<0.73$, it is negative for all $L$ above the positive root of $2a\ln2\,(\tfrac12+a)L^2-2a\ln2\,\delta L-\delta$, which is at most $L^\ast$, where
\[
L^\ast=\frac{\delta}{\frac12+a}+\sqrt{\frac{\delta}{2a\ln2\,(\frac12+a)}}\;\le\;1.5+\frac{1.1}{\sqrt a},
\]
where the last bound uses $\delta<0.73$ and $\tfrac12+a>\tfrac12$. So the cost decreases in $L$ for every $L\ge\max\{L_{\min},1.5+1.1/\sqrt a\}$, for any $\eta$ and any $0<\bar p<p^\ast$. In every row of Table~\ref{tab:llm} we have $L_{\min}\ge6>1.5+1.1/\sqrt a$, so $L\ge L_{\min}$ is the binding condition there.
 
\emph{(d)} At $L=L_{\min}$ we have $2^L\approx B\kappa/L$, so $2^{-2aL}=(L/(B\kappa))^{2a}$. The leading term of (b) becomes
$\tfrac{B\kappa}{L}(1-\eta)^2\big(\tfrac{L}{B\kappa}\big)^{2a}(\tfrac12+a)L=(\tfrac12+a)(1-\eta)^2L^{2a}(B\kappa)^{1-2a}$.
\end{IEEEproof}
 
\begin{IEEEproof}[Proof of Corollary~\ref{cor:tokens}]
(i) Since $n\le q+1$ and $(1-\bar p)^2\le\tfrac12$, we have $n\rho^2\le(2^L+1)(1-\eta)^22^{-L}\le1+2^{-L}<2$. Hence $\lfloor n\rho^2\rfloor\le1$. Under the exact condition $k-1<n\rho^2\le1+2^{-L}$ we get $k\le2$, with $k=2$ only if $n\rho^2>1$.
(ii) Set the right-hand side of \eqref{eq:power} equal to $\lambda_{\rm s}$ and solve for $B$. The choice $L=L_{\min}$ depends only logarithmically on $B$.
(iii) At fixed $B$, the gate enters \eqref{eq:token} only through $\kappa(\tau)$ and $a(\tau)$.
\end{IEEEproof}

\end{document}